\documentclass[aps,prd,preprintnumbers,nofootinbib,twocolumn]{revtex4}
\usepackage{bm}
\usepackage{latexsym}
\usepackage{dcolumn}
\usepackage{amsmath,amsfonts,amssymb}
\usepackage{graphicx,epsfig}
\usepackage{color}
\usepackage{amsthm}

\def\be {\begin{equation}}
\def\ee {\end{equation}}
\def\bea {\begin{eqnarray}}
\def\eea {\end{eqnarray}}
\def\bc {\begin{center}}
\def\ec {\end{center}}
\def\bfg {\begin{figure}}
\def\efg {\end{figure}}
\def\bi {\begin{itemize}}
\def\ei {\end{itemize}}
\def\nn {\nonumber}

\def\la {\label}
\def\le {\left}
\def\ri {\right}

\def\fr {\frac}

\def\b  {\beta}

\def\beq{\begin{equation}}
\def\eeq{\end{equation}}
\def\br{\begin{eqnarray}}
\def\er{\end{eqnarray}}
\newcommand{\eel}[1] {\label{#1}\end{equation}}

\newcommand{\bdm}{\begin{displaymath}}
\newcommand{\edm}{\end{displaymath}}

\begin{document}                         

\title{ On the origin of generalized uncertainty principle from compactified $M5$-brane }

\author{Alireza Sepehri $^{1}$\footnote{alireza.sepehri@uk.ac.ir} }

\affiliation{$^1$ Research Institute for Astronomy and Astrophysics of Maragha
	(RIAAM), Maragha, Iran,P.O.Box:55134-441.}

\author{Anirudh Pradhan $^{2}$\footnote{pradhan.anirudh@gmail.com}}

\affiliation{$^{2}$Department of Mathematics, Institute of Applied Sciences and Humanities, G L A
 University, Mathura-281 406, Uttar Pradesh, India.}
 
\author{A. Beesham $^{3}$\footnote{beeshama@unizulu.ac.za}}

\affiliation{ $^{3}$ Department of Mathematical Sciences, University of 
Zululand, Kwa-Dlangezwa 3886, South Africa.}
\begin{abstract}
In this paper, we demonstrate that compactification in M-theory can lead to a deformation of field theory 
consistent with the generalized uncertainty principle (GUP). We observe that  the matter fields in the 
$M3$-brane action contain higher derivative terms. We demonstrate that such terms can also be constructed 
from a reformulation of the field theory by the GUP. In fact, we will construct the Heisenberg algebra consistent 
with this deformation, and explicitly  demonstrate it to be the Heisenberg algebra obtained from the GUP. Thus, 
we use compactification in $M$-theory to motivate for the existence of the GUP.

\end{abstract}
 \maketitle 
\section{Introduction}
Despite the fact that we as yet do not have a full theory of quantum gravity (QG),
various different approaches to quantum gravity have been proposed. These approaches 
to quantum gravity can be used to gain phenomenological understanding of the nature 
of the actual quantum theory of gravity. It is interesting to note that various different 
approaches to quantum gravity have predicted the existence of a minimum measurable length 
scale. These approaches include string theory and the semi-classical physics of black holes.
The existence of a minimum length deforms the usual uncertainty principle to a (GUP) \cite{q1,q2}

\bea 
\Delta x_i \Delta p_i &\geq& \fr{\hbar}{2} [ 1 + \beta
\le((\Delta p)^2 + <p>^2 \ri) \nn \\
&+& 2\beta \le( \Delta p_i^2 + <p_i>^2\ri) ]~, \la{uncert1}
\eea

where $p^2 = \sum\limits_{j}p_{j}p_{j}$,
$\beta=\beta_0/(M_{p}c)^2=\b_0 \frac{\ell_{p}^2}{\hbar^2}$, $M_{p}=$
Planck mass, and $M_{p} c^2=$ Planck energy. However, as the uncertainty principle 
bears close resemblance to the Heisenberg algebra, the deformation of the uncertainty 
principle also deforms the Heisenberg algebra to \cite{q3,q4}

\be 
[x_i,p_j] = i \hbar ( \delta_{ij} + \beta \delta_{ij} p^2 +
2\beta p_i p_j )~. \la{com1} 
\ee

This form ensures, via the Jacobi identity, that
$[x_i,x_j]=0=[p_i,p_j]$ \cite{q3,q4}.\\

The GUP has opened up an interesting window for quantum gravity phenomenology
because it implies a modification of the fundamental commutation relation
between position and momentum and this in turn also produces correction
terms for all quantum mechanical systems \cite{q12,q14}. It may be noted that 
the GUP deformation has been generalized to include the full spacetime deformation,
and the quantum field theory corresponding to such a deformation has also been 
constructed \cite{w1, w2}.

Even though the GUP has been motivated from various approaches to quantum gravity 
including string theory, it has not been studied within the framework of compactification. 
On the other hand, compactification seems to deform spacetime geometry by a fundamental 
length scale, so it is expected to produce a deformation similar to the GUP deformed field 
theory. In this paper, we demonstrate that  compactification can be used to motivate the deformation 
of matter fields consistent with the GUP. To investigate this, we have to obtain a relevant action 
for the $M3$-brane. 

The gauge field in the action for two M2-branes is known to be valued in a Lie $3$-algebra rather 
than a Lie algebra \cite{q17,q20}. The action for two $M2$-branes can also be constructed without 
using a Lie $3$-algebra as long as the matter fields transform in the bi-fundamental representation. 
However, the Lie $3$-algebra valued gauged fields are important in the construction of a $M5$-brane 
action. This proposal for the $M5$-brane action is based on an infinite dimensional Lie $3$-algebra 
and a Nambu-Poisson structure on three dimensional manifolds. This  model is expected to contain self-dual 
$2$-form gauge fields in six dimensions, and so the resulting action may be interpreted as the action for 
$M5$-brane world-volume \cite{q22}. 

Motivated by these results, we propose a new model for constructing M3-branes from $M0$-branes. This is done 
by replacing the Nambu-Poisson structure in two dimensions by the structure in three dimensions. Then, we 
demonstrate that $N$ $M0$-branes join each other and form a $M5$-brane. This $M5$-brane action is then
compactified with a brane on two circles. Thus, we obtain a DBI action for $M3$-branes. This action contains
extra higher order derivative terms. We observe that these higher order derivative terms are similar
to the terms formed from a GUP deformation of the field theory. In fact, we  will demonstrate that the explicit
form of the modified commutation relations between position and momenta in this model is the same as that produced
by the GUP. It is worth mentioning that the foundations of quantum commutation relations can be motivated from
string theory \cite{Bars:2014jca}.

The outline of our paper is as  follows. In section \ref{o1}, we  construct the M0-brane from the $D0$-brane and 
consider the relation between their algebras. We also show that $M0$-branes can unite and form a $M5$-brane . In 
section \ref{o2}, we will compactify the $M5$-brane on two circles and obtain the relevant action for the $M3$-brane. 
We also demonstrate the relation between the GUP and compactification in this section. The last section is devoted 
to the summary and conclusion.

\section{ $M0$-branes growing into $M5$-branes}\label{o1}

In this section, inspired by the Lagrangian for $M2$-branes and $M5$-branes, we will replace the two dimensional 
Nambu-Poisson bracket by three one in $D0$-branes and build the DBI action for $M0$-branes with a Lie $3$-algebra 
as the internal symmetry. $D0$-branes are $0+1$ dimensional branes in a ten dimensional space-time of string theory 
which has nine transverse scalars corresponding to the nine transverse directions in their action. However, $M0$-branes 
are $0+1$ dimensional branes in the eleven dimensional space-time of M-theory which has ten transverse scalars
corresponding to the ten transverse directions. We will show that these M0-branes can join each other, grow and then 
make a transition to a $M5$-brane. Let us begin with the Born-Infeld action for a $Dp$-brane \cite{p1,pv1,p2,q23,q24,qv24,q25}:

\begin{eqnarray}
S = - T_{p}\int && d^{p+1}\sigma ~ STr \Bigg(-det(P_{ab}[E_{mn} \nn \\ &+& E_{mi}(Q^{-1}+\delta)^{ij}E_{jn}]+ 
\lambda F_{ab})det(Q^{i}_{j})\Bigg)^{1/2}~~
\label{m2}
\end{eqnarray}
where
\begin{eqnarray}
 E_{mn} = G_{mn} + B_{mn}, \qquad  Q^{i}_{j} = \delta^{i}_{j} + i\lambda[X^{j},X^{k}]E_{kj},\label{m3}
\end{eqnarray}
 $\lambda=2\pi l_{s}^{2}$ and $X^{i}$ are scalar fields of dimension mass. Here $a, b=0,1,...,p$
are the world-volume indices of the Dp-branes, $i, j, k = p+1,...,9$
are indices of the transverse space, and $m$, $n$ are the
ten-dimensional space-time indices. Also, $T_{p}=\frac{1}{g_{s}(2\pi)^{p}l_{s}^{p+1}}$
is the tension of $Dp$-brane, $l_{s}$ is the string length and $g_{s}$ is the string coupling.
Using this equation and assuming $G_{ab}=\eta_{ab}+\partial_{a}X^{i}\partial_{b}X^{i}$, we
can approximate Infeld action for $D0$-branes \cite{p1,pv1,q26}

\begin{eqnarray}
S_{D0} =-\frac{1}{2g_{s}l_{s}}\int dt Tr (\Sigma_{i=1}^{9} \{\partial_{t}X^{i}\partial_{t}X^{i}-
\frac{1}{4\pi^{2}l_{s}^{4}}[X^{i},X^{j}]^{2}\})\nn\\
\label{m4}
\end{eqnarray}
where  $X^{i}(i=1,3,...9)$ are transverse scalars.
Following rules for commutation relations, we can write \cite{p1,pv1,p2,q27,q28}

\begin{eqnarray}
&& [X^{0},X^{i}]=i
\lambda \partial_{t}X^{i}\qquad  [X^{m},X^{m}]=0
\label{m5}
\end{eqnarray}
Replacing $\partial_{t}X^{i}$ by $[X^{0},X^{i}]$ in equation (\ref{m5}), we get \cite{p1,pv1,p2}

\begin{eqnarray}
&& S_{D0} =-
\frac{1}{8g_{s}\pi^{2}l_{s}^{5}
}\int dt Tr(
\Sigma_{m=0}^{9}
[X^{m},X^{n}]^{2})
\label{m6}
\end{eqnarray}
Now, from the $D0$-branes, we can construct other $Dp$-branes by substituting the following rules in the action (\ref{m6})
\cite{p1,p2,q27,q28}:

\begin{eqnarray}
&& \Sigma_{m}\rightarrow \frac{1}{(2\pi)^{p}}\int d^{p+1}\sigma \Sigma_{m-p-1} \nonumber \\
&&[X^{a},X^{i}]=i
\lambda \partial_{a}X^{i}\qquad  [X^{a},X^{b}]= i \lambda^{2} F^{ab}\nonumber \\
&& i,j=p+1,..,9\quad a,b=0,1,...p\quad m,n=0,1,..,9~~~~
\label{m7}
\end{eqnarray}
Doing some calculations, we obtain

\begin{eqnarray}
 S_{Dp} =-
T_{p} \int && d^{p+1}\sigma Tr
\Bigg(
\Sigma_{i=p}^{9-p}
\{\partial_{a}X^{i}\partial_{b}X^{i}\nn\\&-&\frac{1}{4\pi^{2}l_{s}^{4}}[X^{i},X^{j}]^{2}+\frac{\lambda^{2}}{4}
(F_{ab})^{2}
\}\Bigg)
\label{m8}
\end{eqnarray}
which is in agreement with studies done for $D1$, $D3$ and $D5$-branes \cite{p1,p2,q24,q26,q27,q28}.
Now, we can extend this mechanism to $M$-Theory and obtain the relevant action for $Mp$-branes by
replacing the two dimensional Nambu-Poisson bracket by three one-dimensional ones in the action (\ref{m6}) \cite{p1,pv1,p2}.

\begin{eqnarray}
S_{M0} =
T_{M0}\int dt Tr(
\Sigma_{M,N,L=0}^{10}
[X^{M},X^{N},X^{L}]^{2})
\label{m9}
\end{eqnarray}

By the compactification of $M$-theory on a circle of radius $R$, this action will
make a transition to the ten dimensional action (\ref{m6}).
To show this, we make use of the method used in \cite{q29},
and define $<X^{8}>=\frac{R}{l_{p}^{3/2}}$,  where $l_{p}$ is the Planck length. Thus, we obtain \cite{p1,pv1,p2}

\begin{eqnarray}
 S_{M0}  &=& T_{M0}\int dt Tr(\Sigma_{M,N,L=0}^{10} [X^{M},X^{N},X^{L}]^{2}) \nonumber \\
&=& 6T_{M0}(\frac{R^{2}}{l_{p}^{3}})\int dt Tr(\Sigma_{M,N=0}^{9}[X^{M},X^{N}]^{2}) + O(X^{6}) \nonumber \\
&=& S_{D0} + O(X^{6})
\label{m10}
\end{eqnarray}
where we define $T_{D0}=6T_{M0}(\frac{R^{2}}{l_{p}^{3}})$. The terms of order  $O(X^{6})$ have not
been written explicitly because they will be ignorable at large radius ($R \rightarrow \infty$). Clearly, 
in this limit, the action (\ref{m10}) for compactified $M$-theory is in agreement with the action (\ref{m6}).

Different $Mp$-branes can be built from $M0$-branes (similar to $Dp$-branes) by using the following rules
\cite{q22}
\begin{eqnarray}
&&[X^{a},X^{b},X^{i}]^{2}=
 \frac{1}{2}(\partial_{a}X^{i})^{2}\qquad  [X^{a},X^{b},X^{c}]^{2}=  (F^{abc})^{2^{â€¢}} \nonumber \\
&&\Sigma_{m}\rightarrow \frac{1}{(2\pi)^{p}}\int d^{p+1}\sigma \Sigma_{m-p-1} \nonumber \\
&&i,j=p+1,..,10\quad a,b=0,1,...p\quad m,n=0,..,10~~
\label{m11}
\end{eqnarray}
where
\begin{eqnarray}
&&F_{abc}=\partial_{a} A_{bc}-\partial_{b} A_{ca}+\partial_{c} A_{ab}\label{m12}
\end{eqnarray}
and $A_{ab}$ is a $2$-form gauge field. We can obtain the relevant action for the $M5$-brane by substituting (\ref{m11}) 
in the action (\ref{m9}):

\begin{eqnarray}
S_{M5} &=&-T_{M5} \int d^{6}\sigma Tr\Bigg(\Sigma_{i=6}^{10}
\{\frac{1}{2}\partial_{a}X^{i}\partial_{b}X^{i}\nn\\&&~~~~~~~~
-\frac{1}{4}[X^{i},X^{j},X^{k}]^{2}+\frac{1}{6}
(F_{abc})^{2}
\}\Bigg).
\label{m13}
\end{eqnarray}
Earlier studies done on the $M5$-branes \cite{q22} agree with the above. The derivation of the action for other $Mp$-branes 
can be done by this mechanism. So, we can also use it to obtain a suitable action for $Dp$-branes by compacting $M$-branes 
on a circle.

\section{Emergence of GUP in compactified $M5$-branes}\label{o2}

In this section, we compactify the $M5$-brane on two circles to obtain $M3$-branes. We do this because our universe has $3 +1$
dimensions, and can be thought to be located on a $3+1$ dimensional brane like a $M3$-brane. In order to do this, we will first 
write the $2$-form gauge fields in $M$-branes in terms of $1$-form gauge fields:

\begin{eqnarray}
&&A_{ab}=\partial_{a} A_{b}-\partial_{b} A_{a}\label{m14}
\end{eqnarray}
For compacting the $M5$-brane to a $M3$-brane, we need to replace two gauge fields by two scalar fields
\cite{p1,p2,q17,q20,q22}. We assume that $A^{4}=X^{4}$ and $A^{5}=X^{5}$. Thus, we obtain

\begin{eqnarray}
&&A_{a4}=\partial_{a} X^{4}-\partial_{4} A_{a}\label{m15}
\end{eqnarray}

\begin{eqnarray}
&&A_{a5}=\partial_{a} X^{5}-\partial_{5} A_{a}\label{m16}
\end{eqnarray}
where $a=0,1,2,3.$ Substituting (\ref{m15}) and (\ref{m16}) in (\ref{m12}), we obtain
\begin{eqnarray}
&&F_{aas}=\partial_{a} A_{as} - \partial_{a} A_{sa}+\partial_{s} A_{aa}\label{m17}
\end{eqnarray}
where $s=4,5.$ We use  the fact that $A_{as}=-A_{sa}$ and $A_{aa}=0$ to obtain

\begin{eqnarray}
&&F_{aa4}=2\partial^{2}_{a} X^{4}- 2\partial_{4} \partial_{a} A_{a} \nonumber \\
&&F_{a44}=2\partial^{2}_{4}A_{a} - 2\partial_{4} \partial_{a}X^{4}\nonumber \\
&& F_{ab4}=2\partial_{a}\partial_{b} X^{4}- 2\partial_{4} \partial_{a} A_{b}
+  2\partial_{4} \partial_{b} A_{a}\label{m18}
\end{eqnarray}
and
\begin{eqnarray}
&&F_{aa5}=2\partial^{2}_{a} X^{5}- 2\partial_{5} \partial_{a} A_{a} \nonumber \\
&&F_{a55}=2\partial^{2}_{5}A_{a} - 2\partial_{5} \partial_{a}X^{5} \nonumber \\
&& F_{ab5}=2\partial_{a}\partial_{b} X^{5}- 2\partial_{5} \partial_{a} A_{b} +  2\partial_{5} \partial_{b} A_{a}
 \label{m19}
\end{eqnarray}
It may be noted that by compacting to a $M3$-brane, the gauge fields do not depend on the fourth and fifth dimensions. 
Thus, we can write
\begin{eqnarray}
&& \partial_{4} A_{a}=0\qquad \partial_{5}  A_{a}=0 \nonumber \\&&\partial_{4} \partial_{a} A_{a}=0\qquad \partial_{5} 
\partial_{a} A_{a}=0\rightarrow \nonumber \\&&
F_{aa4}=2\partial^{2}_{a} X^{4} \qquad F_{aa5}=2\partial^{2}_{a} X^{5}\nonumber \\&& F_{a44}= -2 \partial_{4} \partial_{a}X^{4} 
\qquad  F_{a55}= - 2\partial_{5} \partial_{a}X^{5}\nonumber \\&&
F_{ab4}=2\partial_{a}\partial_{b} X^{4} \qquad F_{aa5}=2\partial_{a}\partial_{b} X^{5}  \label{m20}
\end{eqnarray}
Using the above equations and assuming that the radius of the compactified dimensions are the same, we can calculate the integrals
\begin{eqnarray}
&& x^{4}= x^{5}=2\pi R \rightarrow \nn \\ &&\partial_{4}=\partial_{5}=\frac{1}{2\pi}\partial_{R},\quad \int dx^{4}=\int dx^{5}=
2\pi \int dr \rightarrow\nonumber \\ && \int dx^{4} F_{a44}= -2\int dx^{4} \partial_{4} \partial_{a}X^{4} =\nonumber \\ &&-2\int dx^{5} \partial_{4} 
\partial_{a}X^{4}=2\partial_{a}X^{4} \nonumber \\&&  \int dx^{5} F_{a55}= -2\int dx^{5} \partial_{5} \partial_{a}X^{5} =\nonumber \\ &&-2\int dx^{4} 
\partial_{5} \partial_{a}X^{5}=2\partial_{a}X^{5} \nonumber \\\nn &&\rightarrow \int dx^{4}\int dx^{5}(F_{a44})^{2}=4(\partial_{a}X^{4})^{2} 
\\&& ~~~~~~~~\text{and} \int dx^{4}\int dx^{5}(F_{a55})^{2}=4(\partial_{a}X^{5})^{2}\label{m21}
\end{eqnarray}

Now we can also obtain the rules for commutation as follows
\begin{eqnarray}
&&[X^{4},X^{4},X^{i}]^{2}=
 \frac{1}{2}(\partial_{4}X^{i})^{2}\nn \\ &&[X^{5},X^{5},X^{i}]^{2}= \frac{1}{2}(\partial_{4}X^{i})^{2}
\label{m22}
\end{eqnarray}
Substituting equations (\ref{m20}), (\ref{m21}) and (\ref{m22})
in equation (\ref{m13}), we can obtain the relevant action for the compactified $M3$-brane as
\begin{eqnarray}
 S_{M3} &=&-
T_{M3} \int d^{4}\sigma Tr\Bigg(\Sigma_{i=4}^{10}
\{\frac{1}{2}\partial_{a}X^{i}\partial_{b}X^{i}+4(\partial^{2}_{a} X^{4})^{2}\nonumber
\\&&+4(\partial^{2}_{a} X^{5})^{2}+ 4(\partial_{a}\partial_{b} X^{4})^{2}
+4(\partial_{a}\partial_{b} X^{5})^{2}\nonumber \\&&-\frac{1}{4}[X^{i},X^{j},X^{k}]^{2}+\frac{1}{6}
(F_{abc})^{2}
\}\Bigg)
\label{m23}
\end{eqnarray}
From this action, we can obtain the equations of motion for $X^{4}$ and $X^{5}$:
\begin{eqnarray}
&& \{\partial^{2}_{a} + 4\partial^{4}_{a} + 4\partial^{2}_{a}\partial^{2}_{b}+\frac{\partial^{2} V}{\partial( x^{a})^{2}}\}X^{4}=0\nonumber \\
&& \{\partial^{2}_{a} + 4\partial^{4}_{a} + 4\partial^{2}_{a}\partial^{2}_{b}+\frac{\partial^{2} V}{\partial( x^{a})^{2}}\}X^{5}=0\nonumber \\ &&
V=-\frac{1}{4}[X^{i},X^{j},X^{k}]^{2}
\label{m24}
\end{eqnarray}

Replacing $\partial_{a}=P^{a}$ and $\frac{\partial^{2} V}{\partial( x^{a})^{2}}=m^{2}$, we obtain
\begin{eqnarray}
&& \{(P_{a})^{2} + 4(P_{a})^{4} + 4(P_{a}P_{b})^{2}+m^{2}\}X^{4}=0\nonumber \\
&& \{(P_{a})^{2} + 4(P_{a})^{4} + 4(P_{a}P_{b})^{2}+m^{2}\}X^{5}=0
\label{m25}
\end{eqnarray}
We can compare these equations with the usual equations for scalar field theory

\begin{eqnarray}
&& \{(\tilde{P}_{a})^{2} +m^{2}\}X^{4}=0\qquad \{(\tilde{P}_{a})^{2} +m^{2}\}X^{5}=0
\label{m26}
\end{eqnarray}

As a result, we can redefine the momentum $\tilde{P}$ as

\begin{eqnarray}
&& (\tilde{P}_{a})^{2}=(P_{a})^{2} + 4(P_{a})^{4} + 4(P_{a}P_{b})^{2}\rightarrow
\nonumber \\&& \tilde{P}_{a}\simeq P_{a}(1 +2(P_{a})^{2} +2(P_{b})^{2} ) \label{m27}
\end{eqnarray}

Near the Planck scale, our model modifies the GUP as follows:

\begin{eqnarray}
&&[x_{a},\tilde{P}_{a}]= i\hbar [1+6(P_{a})^{2} +2(P_{b})^{2} ]
\label{m28}
\end{eqnarray}

This equation is  the same as Eq.  (\ref{com1}). Thus, the compactification of a
$M5$-brane action to a $M3$-brane action produces the deformed GUP of Heisenberg. We can argue that
compactification in $M$-theory can also be used as a motivation for GUP. Also, our calculations
show that the GUP occurs only for momentum of scalar fields which are produced from compactification.

\section{Summary and Discussion} \label{sum}
In this paper, we have demonstrated  that all $Dp$-branes can be formed
from joining $D0$-branes. We generalized our model to $M$-theory and showed that
$M5$-branes can also be obtained from $M0$-branes. Then, we compactified an $M5$-brane
on two circles by replacing two gauge fields with two scalars. Thus, we were
able to calculate the relevant action for $M3$-branes. We observed that extra derivatives terms are 
produced due to compactification. These higher derivative terms can be expressed in terms of higher order
momentum terms in the equation of motion for new scalar fields. These terms can be obtained by modifying 
the usual uncertainty principle to the GUP. So, we have demonstrated that compactification in $M$-theory 
can be used as an motivation for the GUP.
\section*{Acknowledgments}
\noindent
The work of Alireza Sepehri has been supported financially by Research
Institute for Astronomy-Astrophysics of Maragha (RIAAM) under research
project NO.1/4717-98. A. Pradhan would also like to thank the University of Zululand, 
Kwa-Dlangezwa 3886, South Africa for providing facilities and support where part of this 
work has been done. The authors are heartily grateful the anonymous referee for his valuable 
comments which have helped to improve and correct the paper in present form.


 \end{document}